\documentclass[aip,app,amsmath,amssymb,reprint]{revtex4-1}
\usepackage{graphicx}
\usepackage{float}
\usepackage{dcolumn}
\usepackage{bm}
\usepackage{tabularx}
\newcolumntype{C}{>{\centering\arraybackslash}X}
\usepackage{multirow}
\usepackage[font={footnotesize}]{caption}
\usepackage[utf8]{inputenc}
\usepackage[hidelinks]{hyperref}

\usepackage{color}
\begin{document}

\title{Realizing $Q > 300,000$ in diamond microdisks for optomechanics  via etch optimization}

\author{Matthew Mitchell}
 \affiliation{Department of Physics and Astronomy and Institute for Quantum Science and Technology, University of Calgary, Calgary, AB, T2N 1N4, Canada}
 \affiliation{Nanotechnology Research Centre, National Research Council of Canada, Edmonton, AB, T6G 2M9, Canada}

\author{David P. Lake}
 \affiliation{Department of Physics and Astronomy and Institute for Quantum Science and Technology, University of Calgary, Calgary, AB, T2N 1N4, Canada}
 \affiliation{Nanotechnology Research Centre, National Research Council of Canada, Edmonton, AB, T6G 2M9, Canada}

\author{Paul E. Barclay}
 \affiliation{Department of Physics and Astronomy and Institute for Quantum Science and Technology, University of Calgary, Calgary, AB, T2N 1N4, Canada}
 \affiliation{Nanotechnology Research Centre, National Research Council of Canada, Edmonton, AB, T6G 2M9, Canada}
 \email{pbarclay@ucalgary.ca}

\date{\today}

\begin{abstract}
Nanophotonic structures in single--crystal diamond (SCD) that simultaneously confine and co-localize photons and phonons are highly desirable for applications in quantum information science and optomechanics. Here we describe an optimized process for etching SCD microdisk structures designed for optomechanics applications. This process allows the optical quality factor, $Q$, of these devices to be enhanced by a factor of 4 over previous demonstrations to $Q \sim 335,000$, which is sufficient to enable sideband resolved coherent cavity optomechanical experiments. Through analysis of optical loss and backscattering rates we find that $Q$ remains limited by surface imperfections. We also  describe a technique for altering microdisk pedestal geometry which could enable reductions in mechanical dissipation.
\end{abstract}

\maketitle

\section{Introduction}

\noindent Over the past decade, advances in fabrication techniques have enabled rapid progress in the development of nanophotonic devices. This maturation is in large part thanks to researchers' ability to borrow techniques and materials from the semiconductor electronics industry. For example, silicon-on-insulator technology, which is designed to reduce stray capacitance of electronic microchips, has provided a ready-made platform for nanophotonic optical  waveguides and cavities capable of confining light to small volumes \cite{ref:thomson2016ros}. In a similar spirit, nanophotonic devices have been realized in a wide variety of dielectric and semiconductor thin films, where the large refractive index contrast between a top thin film waveguide layer and the underlying substrate or removable sacrificial layer provides vertical optical confinement, while patterning of the top layer provides lateral optical confinement. The combination of large per-photon field intensities and low optical loss that can be realized in these devices have enabled studies of strong light--matter interactions that reveal new physical phenomena and can be harnessed for a wide range of applications ranging from sensing \cite{ref:krause2012ahm, ref:wu2014nft, ref:wu2016not}, to nonlinear \cite{ref:evans2013noe, ref:lake2016etv, ref:hausmann2014dnp} and quantum optics \cite{ref:fabre2017qon}.

These innovations in nanophotonics have played a particularly critical role in the field of cavity optomechanics \cite{ref:aspelmeyer2014co}, in which nanofabricated devices are engineered to couple optical and mechanical resonances via optical forces. State-of-the-art cavity optomechanical systems have been fabricated  from dielectric and semiconductor thin films, such as Si \cite{ref:eichenfield2009oc, ref:chan2011lcn}, SiN \cite{ref:eichenfield2009apn, ref:baker2012ois, ref:liu2013eit}, SiO$_2$ \cite{ref:schliesser2009rsc, ref:weis2010oit, ref:park2009rsc, ref:dong2012odm} and III-V semiconductors such as GaAs \cite{ref:ding2010hfg}, InGaP \cite{ref:guha2017hfo}, AlGaAs \cite{ref:guha2017hfo}, GaP \cite{ref:mitchell2014cog}, and AlN \cite{ref:xiong2012ihf}. Owing to the tight optical confinement possible in these thin film based structures, they can posses large optomechanical coupling $g_0$, which quantifies the per-photon optical force in the cavity.  In addition, by undercutting and releasing the waveguiding layer from the substrate via selective chemical etching, they can be mechanically isolated, enabling creation of suspended mechanical resonators with low mechanical dissipation. Together with low optical loss, these properties allow light to be coherently coupled to mechanical resonances of these devices, and operated in the sideband resolved regime commonly used for optomechanical cooling\cite{ref:aspelmeyer2014co}. However, many desirable materials for nanophotonics and optomechanics applications are only available in bulk form, requiring new approaches for fabricating structures with the tight vertical optical confinement and mechanical isolation required by cavity optomechanics.

\begin{figure*}[ht]
  \centering
  \includegraphics[width=1\textwidth]{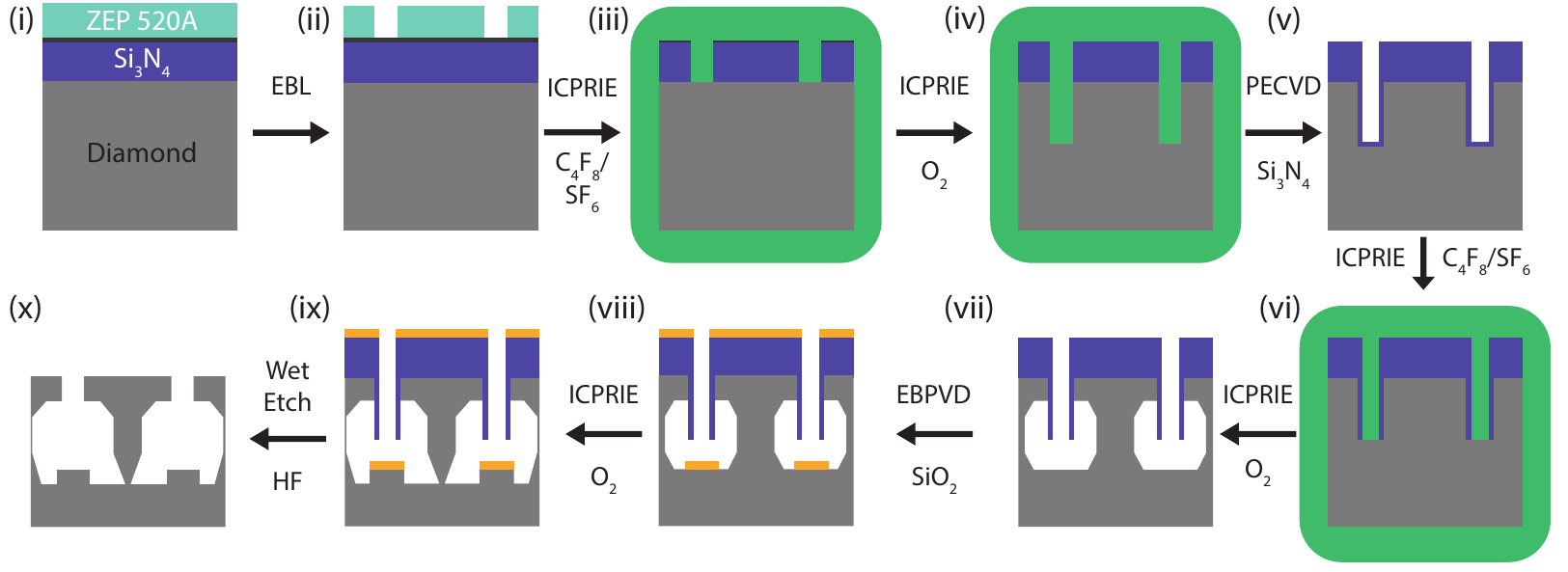}
 \caption{Single--crystal diamond fabrication process, where the steps optimized in this work are highlighted in green. (i) 300 nm thick PECVD Si$_3$N$_4$ layer, $\sim 5$ nm of Ti as an anti--charging layer, and EBL resist (ZEP 520A) is deposited. (ii) EBL is performed and chip is developed in ZED N50 followed by IPA at $-15^\circ$C. (iii) Patterns are transferred to the Si$_3$N$_4$ hard mask using the optimized ICPRIE etch discussed in the text. (iv) ZEP is removed using a deep-UV exposure followed by Remover PG, and the pattern is transferred to the diamond using an anisotropic O2 plasma ICPRIE etch. (v) Sidewall protection layer via a conformal coating of PECVD Si$_3$N$_4$. (vi) A short ICPRIE etch removes Si$_3$N$_4$ from the bottom of the etch windows. (vii) An initial zero bias O$_2$ ICPRIE plasma partially undercuts the microdisks. (viii) A $\sim 100$ nm layer of SiO$_2$ is deposited via electron–-beam physical vapor deposition (EBPVD). (ix) A second zero bias O$_2$ plasma etch is performed, finishing the plasma undercutting process. (x) The sample is soaked in HF to remove the remaining Si$_3$N$_4$ layer, followed by a piranha clean.}
 \label{fig:Process}
\end{figure*}

Single--crystal diamond (SCD) is one such material. Its large electronic bandgap ($\sim 5.45$ eV) and associated wide transparency window results in low multiphoton absorption at telecommunication wavelengths, allowing operation at high optical power levels not supported by smaller bandgap materials such as Si. Among large bandgap materials, diamond has a moderately high refractive index ($n\sim2.45$), allowing for strong optical confinement, as well as excellent mechanical and thermal properties, e.g.\ a Young's modulus and thermal conductivity that are the highest among known materials. Together, these attributes make diamond attractive for nanophotonics and optomechanics applications. Furthermore, diamond is host to quantum emitters such as the nitrogen vacancy (NV) and silicon vacancy (SiV) colour centres \cite{ref:aharonovich2011dp, ref:schroder2016qnd} that can be can be coupled to mechanical \cite{ref:teissier2014scn, ref:ovartchaiyapong2014dsc, ref:lee2016scm, ref:meesala2016esc, ref:meesala2018seo} and optical \cite{ref:englund2010dcs,ref:barclay2011hnr,ref:faraon2011rez,ref:riedrich2014dcs} resonators, and used to generate single photons for quantum networking applications \cite{ref:sipahigil2014ipf, ref:henson2015lfbe,ref:sipahigil2016aid} and store quantum information \cite{ref:balasubramanian2009usc,ref:bar-gill2013ses,ref:sukachev2017svs}.

SCD is not currently commercially available in hetero-epitaxially grown thin film form, and although  efforts to grow high quality diamond films on substrates such as Ir/MgO \cite{ref:washiyama2011cel}, Ir/YSZ/Si \cite{ref:gsell2004art,ref:fischer2008po4}, and SiC/Si\cite{ref:yaita2015hgd} are underway, this material is not yet readily available. Integrated optical and optomechanical devices have been demonstrated in polycrystalline diamond (PCD)\cite{ref:wang2007fac,ref:rath2013wnc,ref:rath2013dio}, which is commercially available in thin film form but is not an ideal host for highly coherent quantum emitters. Most state-of-the-state studies of highly coherent SiV and NV colour centres \cite{ref:balasubramanian2009usc,ref:chu2014cot,ref:evans2016nlh} are  performed using ``bulk'' SCD chips grown using chemical vapour deposition. As such, several alternative approaches to fabrication of SCD nanophotonic devices from this material have been investigated.

Efforts to create nanophotonic devices from SCD include wafer bonding and polishing \cite{ref:hausmann2014dnp, ref:faraon2011rez, ref:lee2012csv,ref:tao2013scd,ref:bayn2014ftn}, liftoff \cite{ref:lee2012csv,ref:piracha2016sfo}, and use of hybrid materials \cite{ref:barclay2011hnr, ref:riedrich2012otd}.  An arguably simpler approach is to fabricate devices directly from bulk diamond chips. This offers the possibility of creating devices from the highest quality material without requiring any manual processing steps such as bonding or polishing.  To this end, ion--beam milling \cite{ref:Babinec2011dfi, ref:hiscocks2008dwf, ref:atikian2017fsn}, angled plasma etching \cite{ref:burek2012fsm, ref:burek2014hqf, ref:burek2016doc}, and plasma undercutting \cite{ref:khanaliloo2015hqv}  approaches have been successful in fabricating nanophotonic structures from SCD.  Of these, the quasi-isotropic plasma undercutting technique developed by Khanaliloo et al.\cite{ref:khanaliloo2015hqv} for fabricating microdisk structures from bulk diamond is unique in its ability to harness etching along diamond crystal planes in both vertical and lateral directions. It has been used to realize diamond devices with a desirable combination of high optical $Q$ and small mode volume \cite{ref:khanaliloo2015hqv} and for the first demonstrations of optomechanics in SCD \cite{ref:khanaliloo2015dnw, ref:mitchell2016scd}.

Recently, a modified version of the undercut process used for early demonstrations of SCD cavity optomechanical microdisks \cite{ref:mitchell2016scd} was used to create devices whose increase in $Q$, together with improvements in their thermal properties, enabled coherent coupling between light and mechanical motion, ie.\ optomechanically induced transparency and cooling\cite{ref:lake2018oit}. In the work presented in this article, we provide a detailed description of this modified fabrication process, and describe additional optimization of the etching parameters that further increases $Q$ by $\sim 4\times$ compared to results reported in Ref.\ \cite{ref:lake2018oit}.  This places the devices in the resolved sideband regime which is a requirement for observing efficient radiation-pressure dynamical back-action effects\cite{ref:kippenberg2008cob, ref:aspelmeyer2014co}, such as ground state cooling\cite{ref:chan2011lcn}, and efficient optomechanical wavelength conversion\cite{ref:hill2012cow, ref:liu2013eit}, which we have recently demonstrated \cite{ref:mitchell2018omw}. In addition, by analyzing the influence of the microdisk's diameter on the loss and internal scattering rates of the microdisk modes, we reveal that $Q$ is likely limited by surface roughness, indicating that further improvements may be possible.

\section{Process overview}\label{sec:Process}

\noindent The approach reported here for fabricating undercut devices from bulk SCD is inspired by the SCREAM process developed for bulk single--crystal silicon MEMS microfabrication \cite{ref:shaw1994sas}, and has been used by our group to fabricate nanophotonic devices such as nanobeams \cite{ref:khanaliloo2015dnw} and microdisk whispering gallery mode resonators \cite{ref:khanaliloo2015hqv}. This process has seen adoption by  researchers fabricating a variety of nanoscale structures from SCD \cite{ref:mouradian2017rpc, ref:wan2018tdp, ref:kiss2018tdg}, and the properties of the quasi-isotropic etch and its interaction with the diamond crystal planes have been investigated in detail in Ref.\ \cite{ref:ling2017cod}.

The modified process used in Ref.\ \cite{ref:lake2018oit}  for fabricating improved microdisk optomechanical cavities is shown in Fig.\ \ref{fig:Process}, where the optimized steps that are the focus of this article have been highlighted in green. We start with 3 mm $\times$ 3 mm optical grade, CVD grown, SCD chips purchased from Element Six\cite{ref:isberg2002hcm,ref:balmer2009cvd}. These are mechanically polished further by Delaware Diamond Knives to surface roughness $<$ 5 nm RMS, and are cleaned in boiling piranha (150 mL H$_2$SO$_4$:50 mL H$_2$O$_2$) followed by a 3 $\times$ 30s rinse in H$_2$O, and drying with N$_2$. As an O$_2$ inductively coupled reactive ion etch (ICPRIE) is used for all of the diamond etching steps it is necessary to use a hard mask for patterning. We have chosen Si$_3$N$_4$ in order to take advantage of highly optimized Si etching recipes during later steps. A 300 nm  thick Si$_3$N$_4$ layer is deposited via plasma enhanced chemical vapor deposition (PECVD). The  Si$_3$N$_4$ is then coated with a thin Ti layer (5 - 10 nm), deposited via electron--beam physical vapor deposition (EBPVD), which reduces charging that can be problematic  during subsequent electron beam lithography (EBL) steps due to diamond's insulating properties. Finally, a 400 nm layer of ZEP 520A is spin coated on the surface of the chip (4000 RPM, 60 s, 180$^\text{o}$C bake for 5 min while semi--covered by reflective lid). Patterning of the ZEP was performed using a Raith 150-TWO system, with beam energy of 30 keV, aperture of 10 $\mu$m, and dose factor of $4\times 80$ $\mu$C/cm$^2$, followed by development for 20 s in a bath of ZED--N50, followed by IPA, both cooled to $-15^\circ$C. The ZEP pattern is then transferred to the Si$_3$N$_4$ hard mask layer via ICPRIE with C$_4$F$_8$/SF$_6$ chemistry using an Oxford PlasmaPro 100 Estrelas Deep Silicon Etching system. This step is vital for ensuring smooth sidewalls of the diamond structures and is discussed in Sec.\ \ref{sec:HardMask}. The pattern is then transferred to the diamond layer via a nearly vertical anisotropic O$_2$ ICPRIE etch, the optimization of which is discussed in Sec.\ \ref{sec:Diamond}.

Before the undercut step, a sidewall protection layer is required to prevent unwanted etching of the patterned device. In this work a 150 nm thick conformal layer of PECVD Si$_3$N$_4$ was used, however, atomic layer deposited Al$_2$O$_3$ has also been successfully employed by Mouradian et al.\cite{ref:mouradian2017rpc}. This layer is then removed only from the top surface and bottom of the etch wells using the same hard mask etch chemistry and conditions as in the patterning step, which preferentially etches horizontal surfaces. The devices are then undercut via a zero bias O$_2$ quasi-isotropic etch performed at an elevated temperature ($250^\circ$C) to reduce the etching time required for sufficient undercutting \cite{ref:mitchell2016scd}. In recent work an additional SiO$_2$ electron beam evaporation step was added to alter microdisk pedestal profiles, as depicted in Fig.\ \ref{fig:Process}(viii-ix)  which led to improved thermal handling, and is discussed in Sec.\ \ref{sec:Diamond}. Finally after sufficient undercutting the hard mask is stripped  in 49\% HF solution (20 mL), and cleaned in boiling piranha, where each acid step is followed  by a 3 $\times$ 30 second rinse in H$_2$O and drying with N$_2$. Further post--processing surface treatments are discussed in Sec.\ \ref{sec:Surface}.

\section{Hard Mask \& Sidewall Protection Layer Etch Optimization}\label{sec:HardMask}

\noindent While EBL offers  ultra--high resolution nano--lithography,  it is vital to make the initial hard mask etch as smooth as possible to reap these benefits. Major limiting factors for $Q$ of previously studied diamond microdisk structures were sidewall roughness and a ``ledge'' on the top edge of the microdisk \cite{ref:mitchell2016scd, ref:lake2018oit}. Both of these imperfections were caused by a non-ideal hard mask etch that impacted the quality of the subsequent vertical diamond etched surfaces.

\begin{figure*}[ht]
  \centering
  \includegraphics[width=\linewidth]{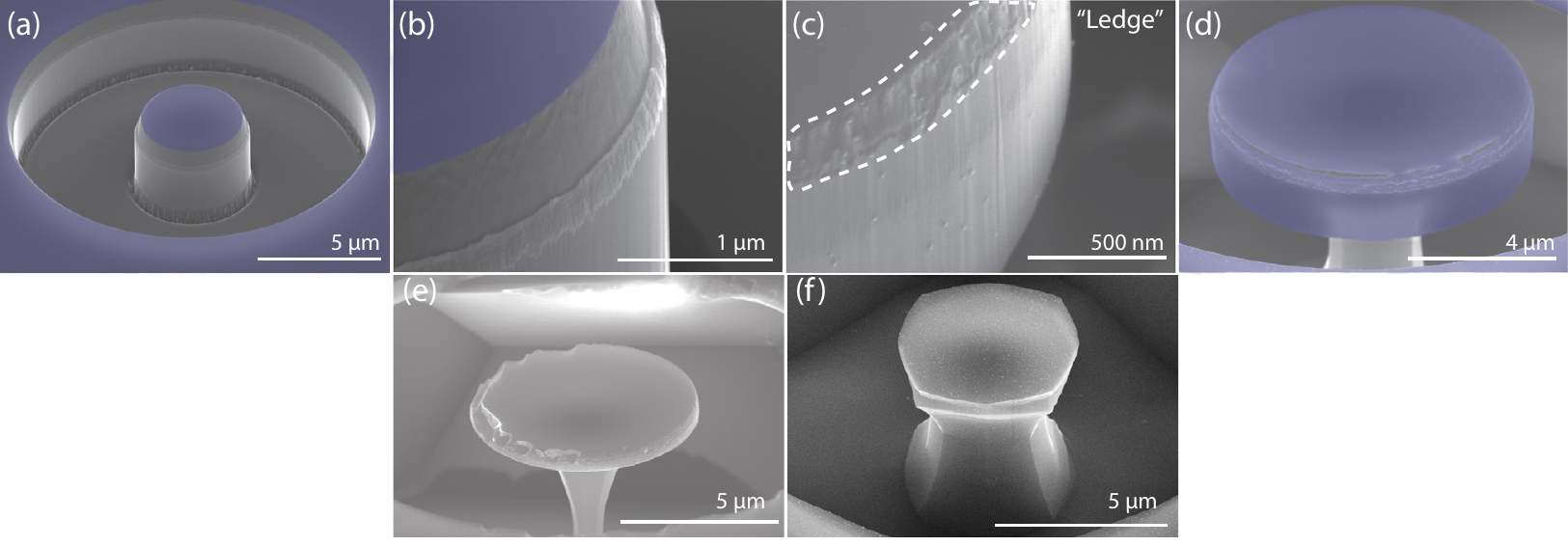}
 \caption{Consequences of poor hard mask etch where the Si$_3$N$_4$ layer has been colorized; non-colorized surfaces are SCD. (a,b) The highly angled sidewalls of the Si$_3$N$_4$ hard mask is a consequence of over--passivation during the etch. (c) This leads to breakthrough of the hard mask layer during the anisotropic diamond etch resulting in ``ledge'' in the diamond layer. (d) More severe breakthrough, resulting in holes in the hard mask, can also result in complete etching of the device. Result of partial (e) and complete (f) breakthrough of the hardmask layer.}
 \label{fig:HardMask1}
\end{figure*}

Fig.\,\ref{fig:HardMask1}(a-c) illustrates the importance of the hard mask etch on the diamond sidewall roughness. The ledge seen in Fig. \,\ref{fig:HardMask1}(c) was a result of breakthrough  of the Si$_3$N$_4$ hard mask's angled sidewalls during the anisotropic diamond etch, as seen in Fig. \,\ref{fig:HardMask1}(d). It was also found that this ledge caused breakthrough of the sidewall protection layer during the quasi--isotropic plasma undercutting step (Fig.\,\ref{fig:Process}(vii)). Prior to the etch optimization described below, this breakthrough severely limited the yield of the process. In extreme cases it resulted in complete delamination of the sidewall protection layer, and subsequent unwanted etching of the device surfaces, as shown in Fig.\,\ref{fig:HardMask1}(e,f).

\begin{figure*}
  \centering
  \includegraphics[width=\linewidth]{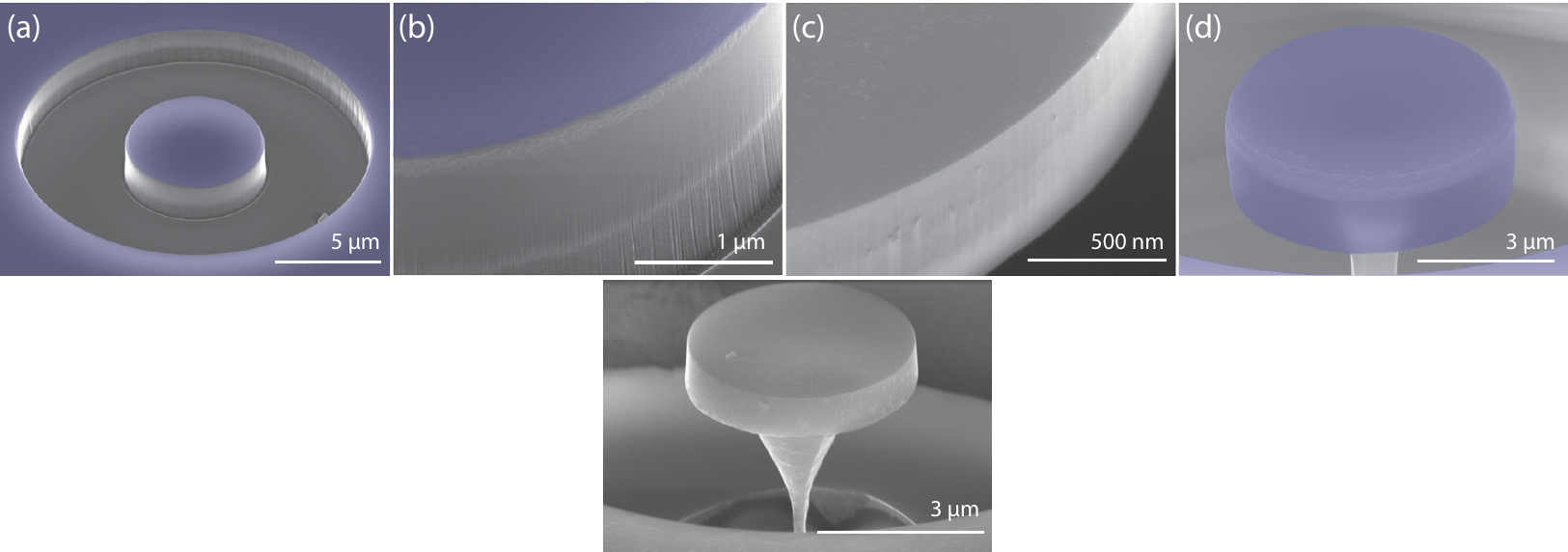}
 \caption{Result of optimized hard mask etch where the Si$_3$N$_4$ layer has been colorized; non-colorized surfaces are SCD. (a,b) The optimized hard mask etch resulted in less angled, smooth sidewalls that do not exhibit breakthrough during the anisotropic diamond etch. (c,d) Smooth diamond sidewalls resulting from optimized hard mask etch, with no holes in the hardmask layer. (e) Example microdisk structure after subsequent processing described in Sec.\ \ref{sec:Process}.}
 \label{fig:HardMask2}
\end{figure*}

The hard mask etch was optimized following a procedure  similar to Hill\cite{ref:hill2013now}, by performing the identical EBL process described above on $\sim 300$ nm thick Si$_3$N$_4$ deposited on Si substrates. This was stopped when the Si$_3$N$_4$ sidewalls were smooth and close to vertical, ensuring that there was no breakthrough during the anisotropic diamond etch as shown in Fig.\, \ref{fig:HardMask2}(a-c). The optimized ICPRIE etch parameters are given in Table\ \ref{tab:SiNEtch}. Here the ICP column refers to the power applied to the coil responsible for generating the plasma, hence controlling the ion density, while the RF column refers to the power applied to the coil responsible for accelerating the ions towards the substrate, controlling ion energy. In optimizing this etch the temperature (15$^\circ$C) and pressure (10 mTorr) were held constant. Initially tests were performed to find a bounding region for the bias voltage, which is a measure of the potential difference between the plasma and substrate electrode, achieved by varying the ICP and RF power and C$_4$F$_8$:SF$_6$ gas ratio. Several iterations were performed holding either the gas ratio or RF and ICP powers constant, while varying the other. The etch quality was analyzed for varying bias voltage and gas ratio until an acceptable etch was achieved, where etch quality was determined by analyzing SEM images post--etching. Overall ratios of  C$_4$F$_8$:SF$_6$ from 3.75 to 0.56, ICP powers from 650 -- 1800 W, and RF powers from 10--25 W were explored.

\begin{table}
\begin{tabularx}{\linewidth}{CCCCCCC}
T & Pressure & RF  & Bias  & ICP  & C$_4$F$_8$  & SF$_6$\\
{[}$ ^\circ$C{]} & [mTorr] & [W] & [V] & [W] & [sccm] & [sccm]\\ \hline \hline
15 & 10 & 20 & 50 & 1200 & 14 & 14 \\
  \hline
\end{tabularx}
\caption{Nominal etch parameters used for patterning the Si$_3$N$_4$ hard mask. This results in an etch rate of $\sim 3.0$ nm/s for PECVD Si$_3$N$_4$.}\label{tab:SiNEtch}
\end{table}

\section{Anisotropic Diamond Etch Optimization}\label{sec:Diamond}

\noindent A smooth vertical diamond etch is also critical for realizing devices with low optical loss. To this end, following the Si$_3$N$_4$ hard mask etch optimization, the anisotropic diamond etching process was optimized to reduce micromasking effects that can lead to rough device sidewalls \cite{ref:madou2011fom}. This optimization is also important for future devices such as photonic crystals, whose sensitivity to roughness is enhanced owing to their large surface area to volume ratio, and whose optical design is simplified if vertical sidewalls are achievable.

\begin{figure}
  \centering
  \includegraphics[width=\linewidth]{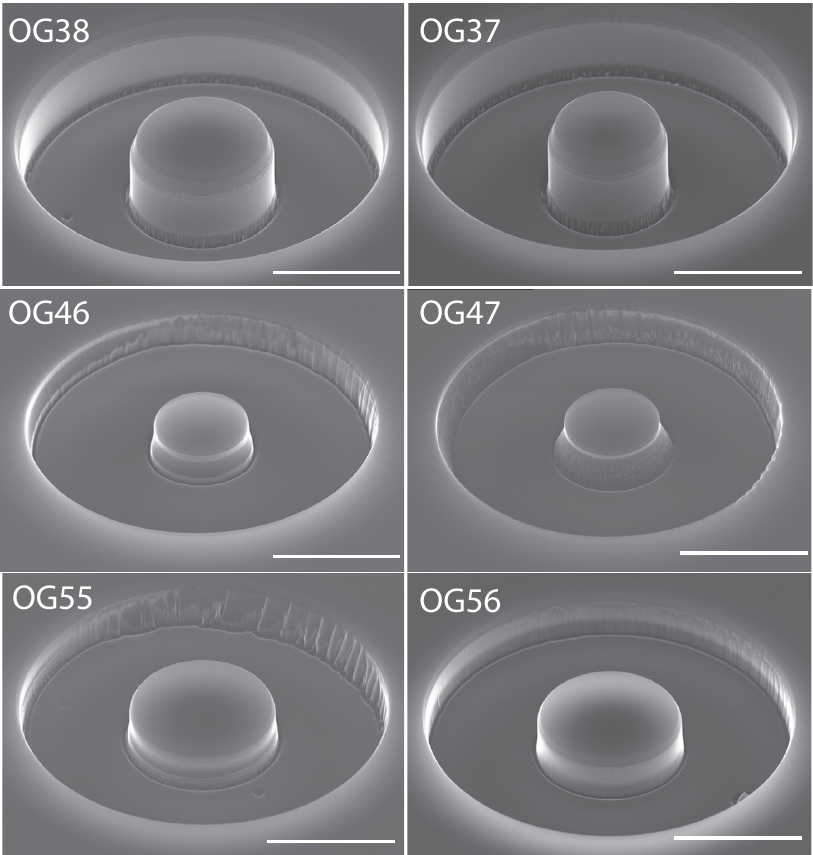}
 \caption{Scanning electron micrographs of microdisk structures after the anisotropic etch step for varying RF power.
 Etch setting at each point in the parameter sweeps RF--$\alpha$ (top two rows of images) and RF--$\beta$ (bottom row of images) are detailed in Table \ \ref{tab:DiamondEtch}.  Scale bars are 5 $\mu$m.}
 \label{fig:RFSweep}
\end{figure}

\begin{figure}
  \centering
  \includegraphics[width=\linewidth]{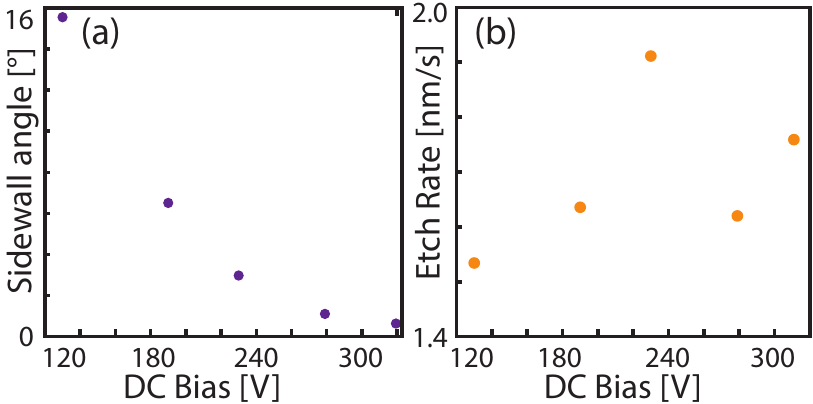}
 \caption{Etch rates and sidewall angles for RF sweep points (RF-$\alpha$) with constant 850 W ICP power. (a) Sidewall etch angle as a function of DC bias, measured from the horizontal. (b) Etch rate as a function of DC bias.}
 \label{fig:RFSweepAngle}
\end{figure}

The anisotropic SCD etches were performed using an Oxford PlasmaPro 100 Cobra ICP system (step (iv) in Fig.\,\ref{fig:Process}), where the goal of this optimization was to create as smooth of diamond sidewalls as possible to reduce optical loss due to surface roughness \cite{ref:borselli2005brs}. During this optimization the temperature (15$^\circ$C), chamber pressure (10 mTorr), and O$_2$ flow rate (30 sccm) of the etches were held constant. Initially a sweep of the etcher RF power from 20 -- 100 W was performed, with the ICP power held constant at 850 W. These etch conditions are labeled RF--$\alpha$ in Table \ref{tab:DiamondEtch}, and SEM images of the resulting etched devices are shown in Fig.\,\ref{fig:RFSweep}. From this sweep it is apparent that low RF power results in highly angled sidewalls and slower etch rates, while the high forward RF power induces some roughness at the bases of the sidewalls. The sidewall angle as a function of DC--bias is shown in Fig.\,\ref{fig:RFSweepAngle}(a), where a maximum angle of $\sim$16$^\circ$ was observed.  The etch rate was found to vary less dramatically throughout this sweep, as shown in
Fig.\,\ref{fig:RFSweepAngle}(b).  Note that roughness in the surrounding etch window walls present in some of these etches is a result of non--ideal EBL and is not a consequence of the individual diamond etch conditions.

\begin{figure}
  \centering
  \includegraphics[width=\linewidth]{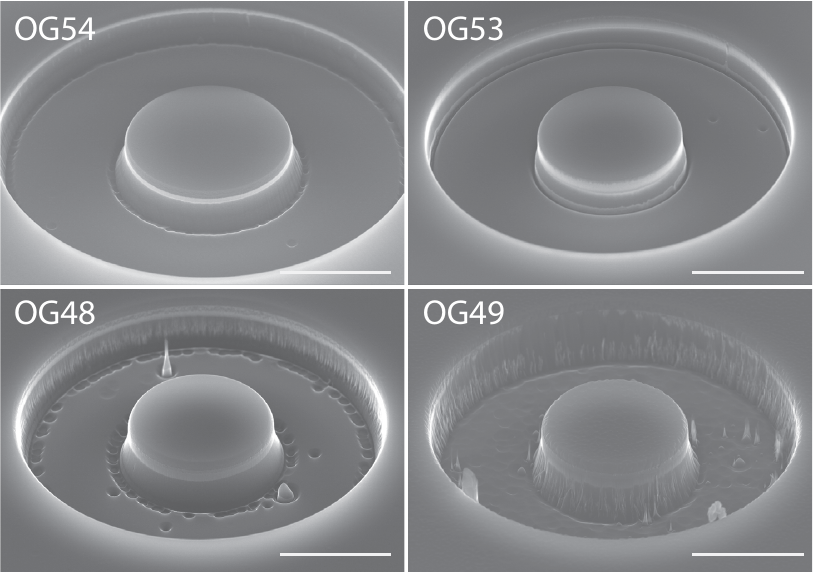}
 \caption{Scanning electron micrographs of microdisk structures after the anisotropic etch step for varying ICP power.  Etch setting at each point in the parameter sweep are detailed in Table \ \ref{tab:DiamondEtch}.  Scale bars are 5 $\mu$m.}
 \label{fig:ICPSweep}
\end{figure}

The smoothest etch, identified to be the OG36 conditions as determined by scanning electron microscope (SEM) images, was then used as a starting point for an ICP power sweep.  During this sweep the DC--bias was kept roughly constant by adjusting the RF power to compensate for variations caused by the changing ICP power. This sweep is labeled  ``ICP'' in Table \ \ref{tab:DiamondEtch}, and SEM images of its results are shown in Fig.\ \ \ref{fig:ICPSweep}.  Based on this sweep, an ICP power of 1000 W was determined to provide the best combination of sidewall smoothness and verticality. This ICP value was used for a final RF sweep, labeled ``RF--$\beta''$ in Table \ \ref{tab:DiamondEtch}. Its results are shown in Fig.\,\ref{fig:RFSweep}, from which the parameters used for sample OG56 were identified as the optimized combination of etch verticality and sidewall smoothness, and were used in the fabrication of the sample whose optical properties are characterized in the following Sec.\ \ref{sec:Surface}.

\begin{table*}
\begin{tabular*}{\textwidth}{c @{\extracolsep{\fill}} cccccc}
 Sample & ICP [W] & RF [W] & Bias [V] & Etch Rate [nm/s] & Sidewall Angle [$^\circ$] & Parameter Sweep \\ \hline \hline
 OG47 & 850 & 20 & 130 & 1.534 & 15.55 & \multirow{5}{*}{RF--$\alpha$} \\
 OG46 & 850 & 40 & 190 & 1.636 & 6.509 &  \\
 OG38 & 850 & 60 & 230 & 1.911 & 2.976 &  \\
 OG36 & 850 & 80 & 279 & 1.620 & 1.107 &  \\
 OG37 & 850 & 100 & 311 & 1.759 & 0.636 & \\ \hline \hline
 OG54 & 850 & 80 & 281 & 2.206 & 13.27 & \multirow{4}{*}{ICP} \\
 OG53 & 1000 & 90 & 286 & 2.454 & 4.063 &  \\
 OG48 & 1150 & 100 & 291 & 3.225 & 6.952 & \\
 OG49 & 1300 & 110 & 293 & 4.062 & 8.994 & \\\hline \hline
 OG55 & 1000 & 100 & 304 & 2.685 & 3.242 & \multirow{2}{*}{RF--$\beta$} \\
 OG56 & 1000 & 110 & 319 & 2.378 & 2.634 & \\
  \hline
\end{tabular*}
\caption{Parameters used in the anisotropic SCD etch optimization.}\label{tab:DiamondEtch}
\end{table*}

\section{Optical Characterization and Surface Treatments}\label{sec:Surface}

\noindent Although the evaluation of the various etch conditions presented above is somewhat qualitative,  a more quantitative evaluation is possible by measuring  the optical properties of a fully fabricated microdisk. This requires further processing steps (vii) -- (x) in Fig.\ \ref{fig:Process} in order to undercut the device, as described in Sec.\ \ref{sec:Process}.  Step (vii) has already been characterized in Ref.\ \cite{ref:khanaliloo2015hqv} and provides a smooth bottom surface for the undercut microdisk structure, vital to observe high-$Q$ optical resonances. The pedestal shaping steps (viii) and (ix) are optional, and are described in Sec.\ \ref{sec:Pedestal} below.  In step (x), after stripping the mask layers the sample was cleaned in heated piranha. During this step the sample was placed in H$_2$SO$_4$ ($6$ mL), heated to 70$^\circ$C before adding H$_2$O$_2$ (2 mL), which raised the temperature of the resulting piranha solution to $\sim 100^\circ$C. After 1 hour the sample was removed and rinsed in H$_2$O (3 $\times$ 30 s) and dried with N$_2$.

\begin{figure}
  \centering
  \includegraphics[width=\linewidth]{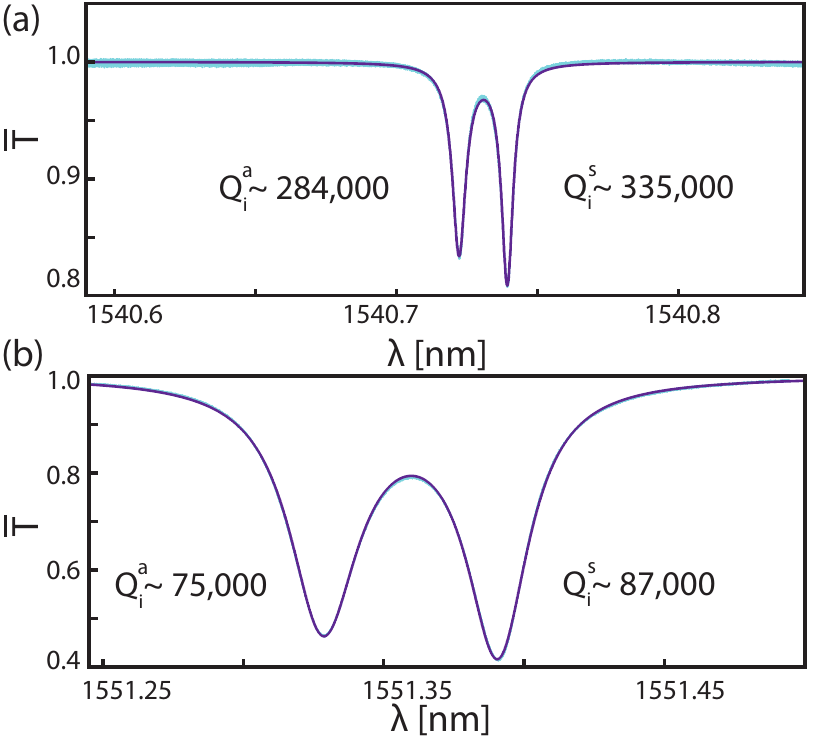}
 \caption{(a,b) Normalized fiber transmission as a function of laser wavelength for an optical mode fabricated using the described optimized process (a) compared to previous work (b), for fixed span in $\lambda$. The intrinsic quality factor $Q_\text{i}$ for the symmetric and antisymmetric modes is extracted by fitting (purple) the transmission profile (blue).}
 \label{fig:HighQ}
\end{figure}

\begin{figure}
  \centering
  \includegraphics[width=\linewidth]{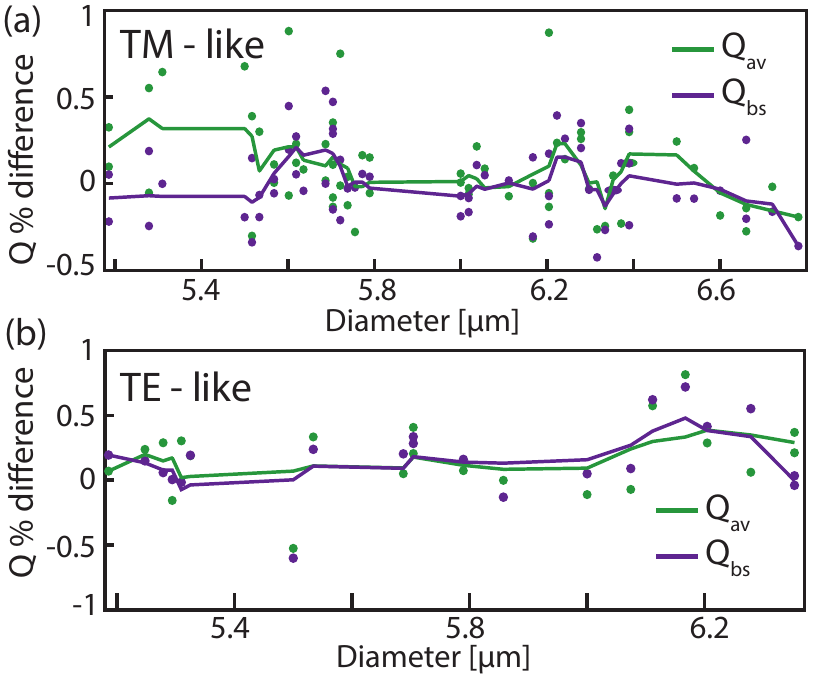}
 \caption{Comparison of $Q_\text{av}$, and $Q_\text{bs}$ before and after cleaning in tri--acid as described in the text for TM--like (a) and TE--like (b) modes.}
 \label{fig:TriAcid}
\end{figure}

\begin{figure*}
  \centering
  \includegraphics[width=\linewidth]{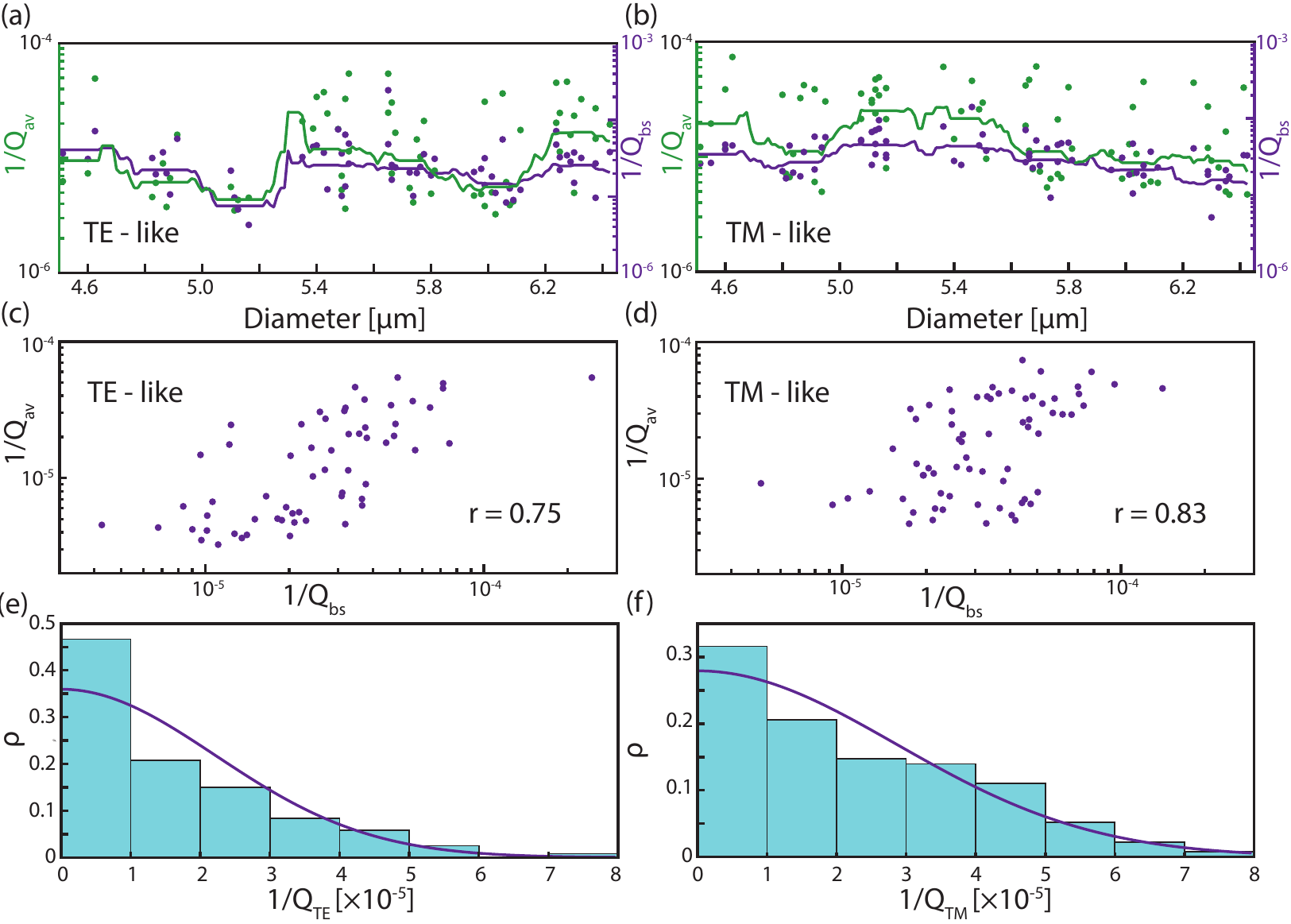}
 \caption{(a,b) $1/Q_\text{av}$ and $1/Q_\text{bs}$ vs. disk diameter for the TE-- and TM--like modes, where an average value of $Q_\text{a}$ and $Q_\text{s}$ is taken. Solid lines are a boxcar average of the data, potted as a guide to the eye. (c,d) $1/Q_\text{av}$ vs. $1/Q_\text{bs}$ for the TE-- and TM--like modes with calculated correlation coefficient, r, suggesting positive correlation between $1/Q_\text{av}$ and $1/Q_\text{bs}$. (e,f) Probability distribution of measured values for (c) $1/Q_\text{TE}$ and (d) $1/Q_\text{TM}$ for microdisks of varying diameter where each histogram has been fit to a half--normal distribution.}
 \label{fig:TETM}
\end{figure*}

The optimized  devices were evaluated by measuring $Q$ of their optical modes and comparing with $Q$ of un-optimized devices. Measurements were carried out by coupling a tunable diode laser (Newport TLB-6700B) to the device via a dimpled optical fiber taper, as outlined in \cite{ref:khanaliloo2015hqv, ref:mitchell2016scd, ref:lake2018oit}. All measurements were performed in a N$_2$ purged environment. Fig.\,\ref{fig:HighQ} compares the fiber taper transmission spectrum for the highest-$Q$ optical mode observed in a optimized fabrication process microdisk (Fig.\ \ref{fig:HighQ}(a)), with that of the highest-$Q$  device from earlier work \cite{ref:lake2018oit} (Fig.\ \ref{fig:HighQ}(b)), demonstrating a $\sim 4\times$ improvement in ``intrinsic'' quality factor, $Q_\text{i}$, to $Q_\text{i} \sim 335,000$. The confidence interval obtained when fitting the optical lineshape to extract $Q$ is typically $\ll 1\%$ of $Q$ and is omitted in the following analysis and figures as it would not be resolved. In this work 67 of 154 pre-- and 88 of 161 post-optimization devices were initially studied as only a subset of the patterned devices are had a sufficiently small pedestal after the undercut to support high--$Q$ modes, as described in Ref. \cite{ref:khanaliloo2015hqv}. From this set only devices possessing a doublet structure, as shown in Fig.\,\ref{fig:HighQ}, were used in our analysis, which corresponds to 62 pre-- and 68 post--optimization devices. This corresponded to the highest--Q modes ($Q > 1.4\times10^4$) of each device, the measurements of which are presented in Figs.\,\ref{fig:TriAcid}-\ref{fig:TETM}. Note that the vast majority of singlet modes observed did not exceed this doublet--$Q$ cutoff. A small handful of singlet modes with $Q$’s up to $5\times10^4$ were observed, but are omitted from this study. The doublet nature of these modes is created by backscattering that couples the microdisk's nominally degenerate clockwise and counterclockwise traveling wave whispering gallery modes to create standing waves \cite{ref:kippenberg2002mct, ref:borselli2005brs}. Each standing wave mode is a symmetric or anti-symmetric superposition of the traveling wave modes, and they can have different $Q^\text{s}_\text{i}$ and $Q^\text{a}_\text{i}$ respectively,  as their intensity profiles sense different volumes of the microdisk.

In general, multiple mechanisms contribute to the total optical loss rate of the microdisks, which scales as $1/Q$. Since loss rates are additive, we can write $1/Q = 1/Q_\text{i} + 1/Q_\text{ex}$. The intrinsic $Q_\text{i}$ can be  decomposed as $1/Q_\text{i} = 1/Q_\text{rad} + 1/Q_\text{b} + 1/Q_\text{ss} + 1/Q_\text{sa}$, where $Q_\text{rad}$, $Q_\text{b}$, $Q_\text{ss}$, and $Q_\text{sa}$ relate to  radiation loss via leakage into unbound modes, bulk absorption, surface scattering, and surface absorption, respectively. Etch smoothness impacts $Q_\text{ss}$, while etch chemistry and other processing that influences the diamond surface can impact $Q_\text{sa}$. $Q_\text{b}$ is determined by the bulk optical properties of the material, and $Q_\text{rad}$ is defined by solutions to Maxwell's equations for modes of a device fabricated without any imperfections. Loss related to coupling with the fiber taper is accounted for the ``external'' $Q_\text{ex}$.

The dominant source of loss can be identified through theoretical calculations or experimental measurements.  For typical fiber taper coupling in these devices, $Q_\text{ex}> 10^6$  is extracted from the fit to the optical resonance. Note that ``parasitic'' loss introduced by the fiber into modes not involved in input or output coupling can also be accounted for as in Spillane et al.\cite{ref:spillane2003ift}, and was found to be small compared to $1/Q$. Bulk loss for a weakly absorbing media can be approximated by $1/Q_\text{b} = \alpha/k$, where the wavenumber $k=2\pi n_r/\lambda = n_r\omega_\text{o}/v_p$ where $\omega_\text{o}$ and $v_p$ are the frequency and phase velocity of the light, respectively. Here $\alpha = 4\pi n_i/\lambda$ is the absorption coefficient of the bulk, where $n_r$ and $n_i$ are the real and imaginary parts of the refractive index \cite{ref:asano2006aot,ref:xu2007nab,ref:kreuzer2008dop}. Using $\alpha \sim 1\times10^{-3}$ cm$^{-1}$ for the absorption coefficient of CVD--SCD at IR wavelengths \cite{ref:mildren2013oed}, results in an estimated $Q_\text{b} > 10^7$. The radiation loss limited contribution was estimated via finite--difference time--domain simulations\cite{ref:oskooi2010mff} and found to be $Q_\text{rad} > 10^6$ for the disk geometry (thickness, radius and pedestal size) studied here.

To investigate loss due to surface absorption, the pre-optimized devices were subjected to tri--acid cleaning. This 1:1:1 sulfuric, perchloric, nitric acid bath  is typically used to remove graphitic surfaces detrimental to  spin coherence properties of diamond nitrogen vacancy (NV) and silicon vacancy (SiV) centers \cite{ref:chu2014cot,ref:evans2016nlh}. It was  performed using 10 mL of the acid mixture at $250^\circ$C for 1 hr using a reflux system to capture the acid vapor, followed by rinsing in H$_2$O (3 $\times$ 30 s) and drying with N$_2$. To assess the impact of this cleaning, microdisk modes of the devices were measured before and after cleaning. These results are summarized in Fig.\,\ref{fig:TriAcid}, which shows the change in measured doublet $Q_\text{av}$, defined as the average of $Q^\text{s}_\text{i}$ and $Q^\text{a}_\text{i}$, for a range devices with varying diameter. This is quantified as ``$Q\%$ difference'', defined as $(Q_\text{after}-Q_\text{before})/Q_\text{before}$ for both $Q_\text{av}$ and $Q_\text{bs}$, for before and after cleaning. No change in $Q_\text{av}$ consistent across many devices or with a clear trend as a function of microdisk diameter is observed. This holds for both TE and TM like mode, whose fields are most strongly concentrated near the vertical and horizontal microdisk surfaces, respectively \cite{ref:borselli2005brs}. Additionally, no significant change in the thermal capacity of the devices was observed, as determined by measuring the shift in doublet center wavelength, $\lambda_\text{o}$, as a function of dropped optical power, $P_\text{d}$.  This suggests that loss due to surface absorption is not limiting $Q_\text{i}$, provided the surface is not contaminated by material impervious to the tri-acid clean, or to the hydrofluoric acid and piranha cleaning carried out post-fabrication (see Fig.\,\ref{fig:Process}). Also shown in Fig.\,\ref{fig:TriAcid} is a measure of the backscattering rate, $Q_\text{bs} \sim \lambda_o/\Delta\lambda$, where $\Delta\lambda$ is the doublet splitting\cite{ref:borselli2005brs}. This was also found to be unaffected by the tri-acid, suggesting that any surface roughness responsible for the mode coupling is likely related to etched diamond surface morphology, which is expected to be unaffected by the cleaning steps used here.

To gain additional insight into the mechanism limiting $Q$, the dependence of $Q_\text{av}$ and $Q_\text{bs}$ on diameter for the optimized devices is plotted in Figs.\ \ref{fig:TETM}(a) and \ref{fig:TETM}(b). Although variations in $Q_\text{av}$ and $Q_\text{bs}$ of an order of magnitude are observed, no clear dependence on diameter is noted, and we attribute the variability to imperfections unique to each microdisk, and to differences in modal radial and vertical quantum number of the modes used in the data set. However, as shown qualitatively in Figs.\ \ref{fig:TETM}(a) and \ref{fig:TETM}(b), we observe that $Q_\text{bs}$ tracks changes in $Q_\text{av}$ as a function of diameter. This is shown more quantitatively in Figs.\ \ref{fig:TETM}(c) and \ref{fig:TETM}(d), which show scatter plots of $1/Q_\text{av}$ as a function $1/Q_\text{bs}$ for the TE and TM modes, respectively. We find that $1/Q_\text{av}$ and $1/Q_\text{bs}$ have a correlation coefficient, $\text{r}=0.83$ and $\text{r}=0.75$ for the TM-- and TE--like modes, respectively. This correlation suggests that surface roughness is limiting $Q$.  Additionally, as shown by the 1/Q histogram in Figs.\ \ref{fig:TETM}(e,f), the density ($\rho$) of high-$Q_\text{av}$ TE--like modes (Fig.\ \ref{fig:TETM}(e)) in these microdisks is larger than that of the TM--like modes (\ref{fig:TETM}(f)). This suggests that there is a greater degree of surface roughness and scattering for the TM--like modes. The probability densities of $1/Q$ for the TE-- and TM-- like modes in Fig.\ \ref{fig:TETM}(e) and  \ref{fig:TETM}(f) were well fit to a half--normal or folded normal distribution expected for $Q$ limited by surface roughness that varies randomly along the perimeter of the microdisk, with different random distribution for each disk. Finally, no large asymmetry between $Q^\text{s}_\text{i}$ and $Q^\text{a}_\text{i}$ of  the standing wave modes was observed for any of the devices measured in this study, further supporting the conclusion that the dominant form of surface roughness limiting $Q$ is distributed along the perimeter of the microdisk. This is in contrast to the effect of large discrete local perturbations along the microdisk perimeter, which can couple differently to the phase shifted spatial intensity profiles of the standing wave modes \cite{ref:borselli2005brs}.

Based on the evaluation of each loss mechanism described above we conclude that $Q_\text{ss}$ is limiting $Q_\text{i}$, most likely due to remaining etch roughness in the anisotropic diamond etch and roughness on the top or bottom surfaces. This suggests that further improvements to $Q$ could be achieved by developing a smoother anisotropic diamond etch.

\section{Thermal Engineering via Pedestal Shape}\label{sec:Pedestal}

\begin{figure*}
  \centering
  \includegraphics[width=\linewidth]{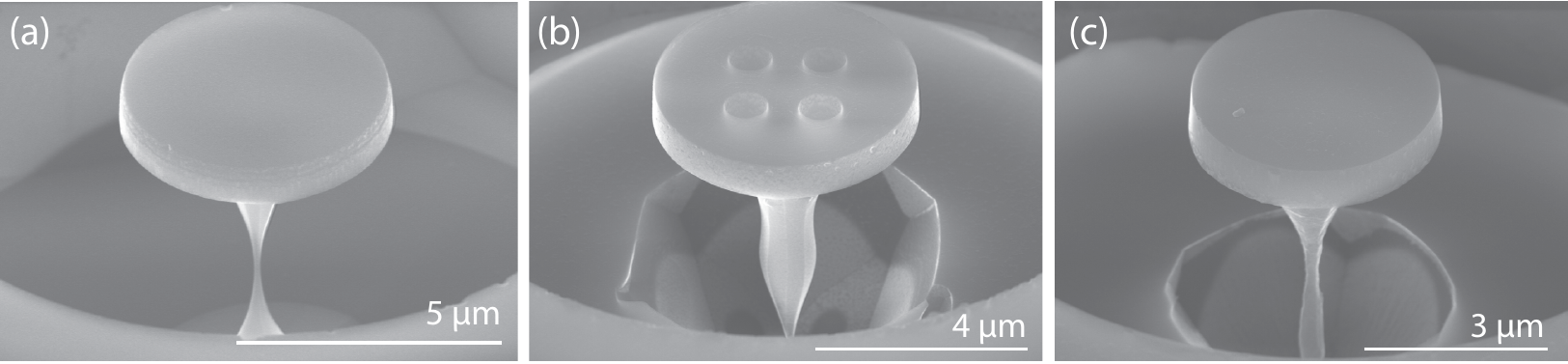}
 \caption{Effect of additional masking step during the quasi--isotropic undercut on microdisk pedestal geometry. (a) Hourglass pedestal shape resulting from no additional masking step. (b) Flared shape resulting from masking after undercutting by $\sim 45\%$. The circular holes observed in the microdisk were part of a separate study, and did not affect the flared pedestal shape. (c) Pedestal shape resulting from masking after undercutting by $\sim 40 \%$.}
 \label{fig:Pedestal}
\end{figure*}

\noindent When confining light to a small mode volume in a solid state structure both linear and nonlinear absorption of light can cause significant heating of the cavity \cite{ref:carmon2004dtb,ref:barclay2005nrs}. This heating can cause instability in the cavity resonance frequency due to the thermo--optic effect and make it practically difficult to maintain a constant--cavity detuning, which is vital in many optomechanical processes. In previous work it was found that the $\sim 100$ nm diameter pedestal size of the microdisk structure pictured in Fig.\ \ref{fig:Pedestal}(a) was limiting the thermal time constant of the device due to the reduced thermal conductivity in the pedestal \cite{ref:mitchell2016scd}.

To attempt to alleviate this issue a method for altering the pedestal shape was investigated. Namely by depositing an additional masking layer it is possible to tailor the pedestal shape of the microdisk during the undercut stage. It is important to use a non-conformal layer such that the undercutting process may continue horizontally as shown in Fig.\,\ref{fig:Process}(viii). Here EBPVD was used to deposit a 100 nm layer of SiO$_2$, in a line of sight fashion, allowing undercutting to continue immediately after. Here a Johnson Ultravac load-locked electron beam evaporation system was used with a deposition pressure of $\sim 1\times 10^{-6}$ Torr, and a deposition rate of 0.5 \AA/s. By varying the point at which this deposition occurs during the undercut the pedestal shape can be altered differently as shown in Figs.\ \ref{fig:Pedestal}(a--c). Here the SiO$_2$ was deposited after 4 hours of undercutting for the structures in Figs.\ \ref{fig:Pedestal}(b,c), where an additional 4 and 5 hours were required to undercut each device respectively. This corresponds to 45\% and 40\% of the total undercut time, where the discrepancy in time is due to different etch depths during the anisotropic etch\cite{ref:khanaliloo2015hqv}. This process can be repeated to customize the pedestal shape and could potentially be used to engineer the phononic properties of the structure, similar to the work of Nguyen et al.\cite{ref:nguyen2015iod}. By altering the pedestal geometry, the device in Fig.\ \ref{fig:Pedestal}(b) demonstrated an order of magnitude decrease in the thermal time constant of the microdisk structures, measured by fitting the response of the optical transmission for an input optical step function. This alteration resulted in the ability to support roughly an order of magnitude larger intracavity photon number before the onset of thermal instability \cite{ref:lake2018oit}. The ability to operate at large $N$ while avoiding thermal instability is practically important in cavity optomechanics applications due to the linear dependance of the optomechanical cooperativity $C=4g_0N/\kappa\Gamma_\text{m}$ on $N$, where $g_0$ is the single--photon coupling rate, $\kappa = \omega_\text{o}/Q_\text{o}$, and $\Gamma_\text{m} = \omega_\text{m}/Q_\text{m}$ \cite{ref:aspelmeyer2014co}. This enabled the observation of optomechanically induced transparency and optomechanically mediated wavelength conversion with $C>1$ in previously reported work\cite{ref:lake2018oit,ref:mitchell2018omw}.

\section{Discussion}

\noindent

The detailed description of the fabrication process provided here will enable researchers to  create a wide range of diamond  photonic devices for applications including quantum photonics, nonlinear optics, and optomechanics. The optimization presented here, resulting in a $\sim4\times$ increase in average $Q_\text{i}$ for TE--like modes compared to our previous cavity optomechanical devices, was carried out over a timescale of $\sim$3 months on equipment shared with other researchers for a wide variety of processes and materials.  As etch parameters vary from tool to tool, we hope that the optimization results presented here could be utilized by others to identify similar issues or devices characteristics and use our procedure to improve overall etch quality. For reference, Table\ \ref{tab:comp} compares this result with the current state of the art for SCD cavities at telecommunications wavelengths. While SCD optomechanical crystals, demonstrated by Burek et al.\cite{ref:burek2016doc} provide superior $g_0$, microdisks have an advantage in that they naturally support optical modes across the entire transparency window of the material, enabling multimode optomechanical experiments such as optomechanical wavelength conversion\cite{ref:hill2012cow,ref:liu2013eit,ref:mitchell2018omw}, and for larger microdisk diameter can have much greater radiation loss limited optical $Q$. The improvement demonstrated here is particularly meaningful for applications in cavity optomechanics, as it places the optimized devices in the resolved sideband regime where the mechanical resonance frequency, $\omega_\text{m}/2\pi \sim 2 - 3$ GHz for the microdisks studied here, exceeds the cavity optical energy decay rate, $\kappa/2\pi \sim 0.6$ GHz for the high--$Q$ device \cite{ref:aspelmeyer2014co}. This regime is a requirement for observing efficient radiation-pressure dynamical back-action effects\cite{ref:kippenberg2008cob,ref:aspelmeyer2014co}, such as ground state cooling\cite{ref:chan2011lcn}, and low--noise amplification \cite{ref:ockeloen-korppi2016lna}. Further enhancement in optical $Q$ could be achieved by improving the etch quality, and reducing surface roughness that is still present, as we conclude from the analysis above that the optical $Q$ remains limited by surface imperfections.  For example, incorporation of Cl$_2$ based etching may enable smoother diamond surfaces \cite{ref:lee2008emo}. Additional cleaning steps such as post--fabrication oxygen annealing, as utilized by Burek et al.\cite{ref:burek2016doc}, or investigating appropriate surface termination techniques such as those devoped for silicon\cite{ref:borselli2006mrs} could also be investigated for improving optical $Q$.  Finally, while current SCD microdisks mechanical quality factors, $Q_\text{m}$, are limited by clamping loss \cite{ref:mitchell2016scd,ref:lake2018oit}, the pedestal shaping step described in Sec.\ \ref{sec:Pedestal} could be utilized to reduce mechanical dissipation and increase $C \gg 1$, by incorporating a phononic shield into the pedestal as demonstrated by Nguyen et al.\ in GaAs microdisks \cite{ref:nguyen2015iod}.

\begin{table}
\caption{Comparison of SCD cavities supporting modes at telecommunications wavelengths. Indicates which works have also demonstrated optomechanical coupling in the structure.}
\begin{tabular}{llcc}
  \hline
    Author/Reference & Structure & $Q_\text{i} (\times 10^5)$ & Optomechanics  \\ \hline
    This work & Microdisk & 3.35 & Yes  \\
    Lake et al. \cite{ref:lake2018oit} & Microdisk & 0.87 & Yes \\
    Mitchell et al. \cite{ref:mitchell2016scd} & Microdisk & 0.68 & Yes  \\
    Khanaliloo et al. \cite{ref:khanaliloo2015hqv} & Microdisk & 1.02 & No  \\
    Burek et al. \cite{ref:burek2014hqf} & Racetrack & 2.70 & No \\
    Burek et al. \cite{ref:burek2016doc} & OMC & 3.02 & Yes \\
    Teodoro et al. \cite{ref:teodoro2018scd} & Microdisk & 0.057 & No \\
  \hline
\end{tabular}
\label{tab:comp}
\end{table}

\section{Acknowledgements}
\noindent The authors would like to thank Blaine McLaughlin for his assistance in the lab, and Ghazal Hajisalem, Aaron Hryciw, Scott Munro and Les Schowalter for all their help and support in and out of the NRC Nanotechnology Research Centre and nanoFAB.

\section{Funding}
\noindent This work was supported by National Research Council Canada (NRC), Alberta Innovates, National Sciences and Engineering Research Council of Canada (NSERC), and Canada Foundation for Innovation (CFI).

%

\end{document}